# Location Optimization of ATM Networks

Somnath Basu Roy Chowdhury
Indian Institute of Technology
Kharagpur, West Bengal 721302
brcsomnath@ee.iitkgp.ernet.in

Biswarup Bhattacharya
Indian Institute of Technology
Kharagpur, West Bengal 721302
biswarup@iitkgp.ac.in

Sumit Agarwal
Indian Institute of Technology
Kharagpur, West Bengal 721302
sumit.agarwal@iitkgp.ac.in

## ABSTRACT
ATMs enable the public to perform financial transactions. Banks try to strategically position their ATMs in order to maximize transactions and revenue. In this paper, we introduce a model which provides a score to an ATM location, which serves as an indicator of its relative likelihood of transactions. In order to efficiently capture the spatially dynamic features, we utilize two concurrent prediction models: the local model which encodes the spatial variance by considering highly energetic features in a given location, and the global model which enforces the dominant trends in the entire data and serves as a feedback to the local model to prevent overfitting. The major challenge in learning the model parameters is the lack of an objective function. The model is trained using a synthetic objective function using the dominant features returned from the k-means clustering algorithm in the local model. The results obtained from the energetic features using the models are encouraging.

## 1 INTRODUCTION
An interbank network, also known as an ATM consortium or ATM network [6], is a computer network that enables ATM cards issued by a financial institution that is a member of the network to be used to perform ATM transactions through ATMs that belong to another member of the network. Interbank ATMs are particularly useful in cases the ATM branch of the same bank is not nearby and more importantly they allow customers to have surcharge (a fee deducted by the banks on transactions from non-bank ATM cards) free transactions in the same network.

American Chartered Bank, U.S. Bank, Capital One 360, Whitney Bank, 1st Advantage Federal Credit Union, Workers Credit Union are examples of some banks which have ATM networks all over the United States. These ATM networks have also collaborated with bigger organizations like 7-Eleven [10], WalMart etc. to install ATMs at their stores to enhance customers convenience [2]. Hence, comparing their competitive strength is a motivating task in order to analyze who has a strategical edge over others in the market.

The ATM networks always try to maximize their return on investment by planning strategically their ATM network so as to maximize the number of transactions [7]. The various factors that a bank keeps account of while placing an ATM include, but is not limited to, popularity of location, type of customer traffic, intensity of traffic, approximate number and type of transactions, service routes (especially important for safety purposes), ATMs of competition in the vicinity, leasing space availability and costs, utilities & infrastructure, potential criminal activity, maintenance & power costs etc. The companies also consider the cost of setting up an ATM in such locations and the amount of return that can be estimated. At last the companies can analyze whether the entire venture is successful.

We define a similar problem where we consider the ATM locations of different banks in the state of California, USA. The location features like the population density, average income, living standard can be gathered from the given zipcodes. We try to form a logical inference of the features at hand in order to assign accurate priority to the appropriate features. After the assignment, we sort regions where maximum transactions are possible and form an inference about the regions where placing ATMs will provide maximum revenue.

## 2 METHODOLOGY
Our entire methodology is divided into the following sections: data pre-processing, visualization of extracted features, inferring the priority weights to be assigned to each features and deduction. Post this, we exploit the weights to fit a regression to compute revenues generated by each ATM.

### 2.1 Dataset Description & Preprocessing
Our dataset is constructed from the publicly available data of ATM locations in the state of California, USA. The original dataset consists of only 3 features viz. city name, street address and zipcode. The dataset consisted of 11229 ATM locations (not necessarily unique zipcode). From the zipcode and street address available, dataset expansion was done from a website which listed US zip codes [9]. For every zipcode 73 more features were extracted by crawling the website. Several other features were also added in order to complete the dataset as described in the subsequent subsections.

(1) *Feature Extraction*: In order to come up with a holistic prediction model based on ATM location, we have collected several features from [9].
(2) *Latitude & Longitude Extraction*: The latitude and longitude of individual zipcode is crawled using a python library, pygeocoder 1.2.5 [11], which provides a convenient API for this task.
(3) *Frequency of ATMs*: The number of ATMs in a particular zipcode is calculated. This depicts the ATM density in a particular region. Various other inferences can be drawn regarding the advantage a particular ATM has over others depending on the population density in the area, and the overall status of people residing in the area.

(4) *Labeling the ATM name tag*: This feature forms an integral part of our overall estimation model. For every ATM location based on the street address it was classified into one of the 7 categories in the Table 1. A relative score was assigned to each class. This score is provided manually based on the relative frequency of daily visitors in the location.

Table 1: Relative Scores for Name-tag class

| Name-tag class | Relative Score |
| --- | --- |
| Shopping Malls | 10 |
| Banks/ Exchange Centre | 9 |
| Recreation Centre | 8 |
| Gas Stations/Car wash | 7 |
| Office Area | 6 |
| Individual Store | 5 |
| Null Data | 4 |

(5) *Nearest zipcode computation*: Data relating to the nearest zipcode sorted by distance was also collected from the United States zipcode website [9]. The nearest zipcode provides insight into the likelihood of people using an ATM in a nearby region.

## 3 APPROACH

The estimation process involves assigning a given weight to every zipcode ($w_{zip}$) which is used to compute the score of an ATM location ($S_{name-tag} \times w_{zip}$), where $S_{name-tag}$ is the relative score based on name-tag. The learning of the zipcode weights is a challenging task as we do not have a loss-function on which it can be optimized.

In order to overcome the complication of having an objective function, we attempt to unearth the relation between various features and revenue generation by the ATMs for each county. In this particular example, to account for simplicity of our model, we assume that the cost per transaction is uniform across all the ATM networks. Using this, we can reduce the problem to finding the total number of transactions for each county. A suitable objective function which denotes the *Wealth Estimate* is chosen. The choice of such this objective function is justified from the results of the local model.

### 3.1 Wealth Estimate

Since the data pertaining to number of transactions is not explicitly provided to us, we proceed by computing another measure, called the Wealth Estimate (WE) for each zip code, which is related as follows:

$$WE = PD \times MHI \times (1 - PNE) \quad (1)$$

where, *PD* refers to the normalized population density of the zip code, *MHI* refers to the normalized median household income of the people residing in the zipcode and *PNE* refers to the percentage of people not earning in the zipcode.

The visualizations rendered henceforth utilize the assumption of the objective function and attempt to find relations between several demographic features with this metric. Thus, the next section deals with the association of some relevant features with the Wealth Estimate, and the inferences we can draw from them.

## 4 ILLUSTRATIONS

Experiments involving the features collected were conducted in order to construct the ideal prediction model. The various features which hold a high correlation coefficient with our objective function are listed below. The plots are obtained using the software Tableau 10.2 [3].

(1) *Rented 1-Bedroom Houses*: Figure 1(b) demonstrates the correlation between the Wealth Estimate, and the percentage of occupied houses which have only one bedroom. The calculated correlation coefficient is 0.467, which presents a positive correlation between the Wealth Estimate and the above variable.
(2) *Median Home Value*: Figure 1(a) demonstrates the correlation between the Wealth Estimate and median home value of the area represented by the zip code. The calculated linear correlation coefficient is 0.45.
(3) *Transport*: The correlation between the Wealth Estimate, and the cumulative percentages of people(above 16 years) who commute via public transportation modes and using cabs and motorcycles versus the wealth estimate as shown in Figure 1(c). The observed linear correlation coefficient is 0.6088.
(4) *Private Primary Schools*: Those areas which had a greater percentage of students in the age bracket of 3-17 years enrolled in private primary schools were observed to have higher wealth estimates, again alluding and augmenting the overall economic status of the area is shown in Figure 1(d). The calculated linear correlation coefficient is 0.41.

The graphs for the above features has been shown in Figure 2. A total of 11 features were considered for the global objective function. The features along with their corresponding weights are: Transportation (0.13), Employment (0.083), Percentage in Private Primary Schools (0.083), Median Home Value (0.093), Percentage of Rented People having single bedroom (0.102), Educated Section (0.074), Population Density (0.167), Median Home Income (0.045), Percentage of people earning (0.083), Percentage of people who are single (0.065), Percentage of single people with roommates (0.074).

## 5 MODEL

In order to analyze accurately the variation of the number of ATM transactions we emphasize on the observation that the dominant features for the entire state California may not affect a particular county with the same impact. For this reason, two separate models are deployed: one for capturing the global features and another for the local features. Both the models are used to compute the weights for the features and the predictions are computed as shown in the Figure 2.

### 5.1 Global Model

In order to formulate the global model, the entire dataset is considered and the features mentioned in the section 4 are taken into account. The weights are computed using the softmax regression shown in Equation 2. After computing these weights, the corresponding features $X$ are considered and a linear combination was applied as shown in Equation 3 to compute the weight for each zip-code ($y_{global}$).

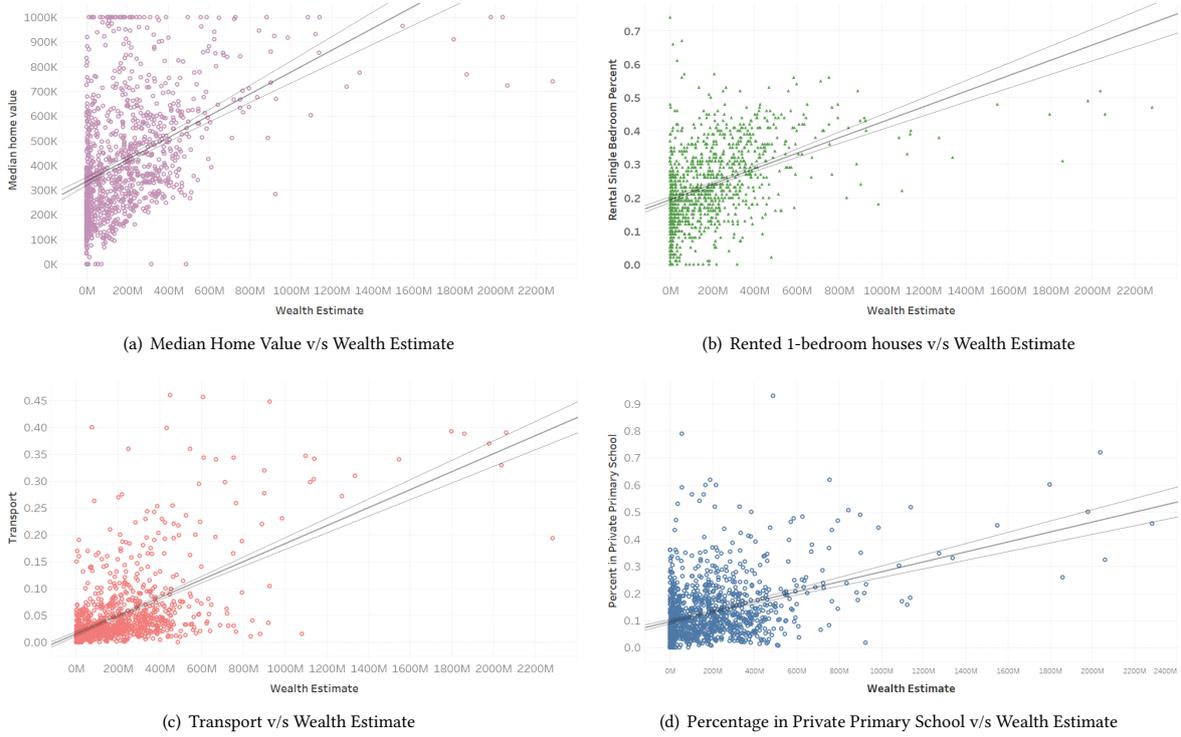

Figure 1: Illustrations for various features v/s Wealth Estimate

$$w_{i,norm} = \frac{exp(w_i)}{\sum_{i=1}^{n} exp(w_i)} \quad (2)$$

$$y_{global} = w_{norm}^T X \quad (3)$$

The relevance of the above features have been discussed in the previous section. The score $y_{global}$ is combined with $y_{local}$ to produce the final score. This will be discussed in our next subsection.

### 5.2 Local Model

In order to shed more light on the local features of every zip-code, we designed the local model. There would be some features that govern ATM revenue generation on a county level, and these may be more abstract than the granular features that the zip-code level analysis has to offer. For analyzing this, we partitioned the data by county, thus generating a number of data-sections (equal to the number of counties). For each of the data sections, we applied a separate k-means clustering algorithm [5] to group similarly behaving features closer to each other in high dimensional space. After some experimentation and trade-off between computation and relevance, we settled on a $k = 7$. Consequently, this resulted in allotting 7 labels $y$ to each of the zip code based samples for each county. To reverse engineer the feature importance, the output labels, $y$, thus generated are used as a class label for supervised classification using a random forest classifier [1]. This enabled us to reveal the relevance of each feature and hence interpret the feature importance as the weight $w_i$. The weights could further be normalized and exploited as exactly as in Equations 2 and 3 respectively. The k-means clustering method is an iterative clustering algorithm, relies on the convergence of data points to group themselves in $k$ clusters, which are decided by the nature of the data. The idea to be conveyed is that similar data points will occupy positions which are closer to each other in the high dimensional space. The random forest classifier[1] is an ensemble learning technique which uses multiple weak learners, which in this case are decision trees to come to a consensus about the output, which in this case is the output label fitted by the clustering algorithm.

After determining the weight vectors, the most energetic features were chosen, top 20 according to the decreasing order of their weights. The linear combination is computed only using these 20 features, as they are relevant to our local problem. Note that for each county, the above computation is conducted, and the top 20 features chosen are not the same every time. Each of the features thus selected demonstrate characteristics specific to the county. Analyzing those features, we try to find the most common features among the pool of features. This provides us with an important conclusion about the function that we assumed as our objective function. Table 2 shows that our initial assumption of the objective was correct and the three factors considered are among the top 10 attributes. The local model maps the most important features for a particular county. We can conclude that our initial function (Equation 1) was reasonable as all the factors are positively correlated and are among the top influential factors.

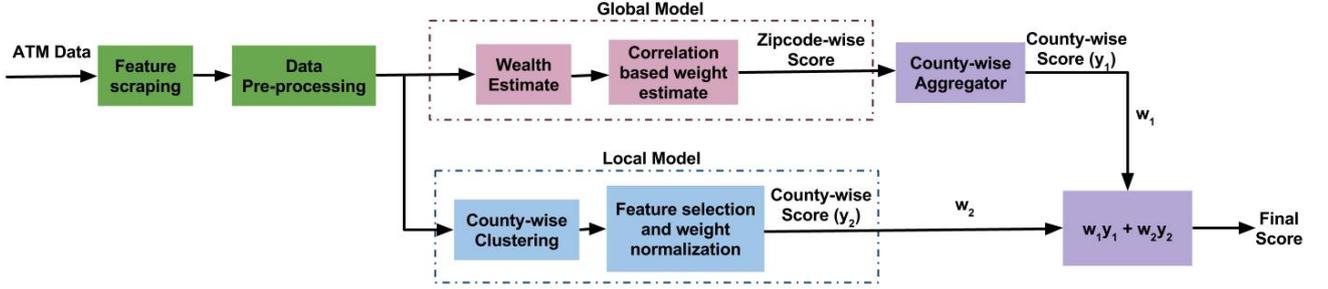

Figure 2: Complete Workflow of the Estimation Model

Table 2: Frequency of Features in Objective Function

| Attribute | Rank |
|---|---|
| Population Density | 1 |
| Employment Status | 7 |
| Median Household Income | 10 |

## 6 RESULTS

From the observation in the visualization section we assign weights to the respective features. The weights are crucial in assigning the score of each zip-code in the data-set. We consider the zip-codes as the building blocks and then sum them up county-wise to form the strategic advantage.

### 6.1 Inferences from visualization

From the global model we retrieve a set of 11 features which according to our model should affect the number of ATM transactions in a locality. In this section, we discuss the intuitive implications and relevance of the energetic features. The most energetic features: Population density and Transportation, both signify that positioning an ATM in a populated area is advantageous. The positive correlation of the features Median Home Value, Percentage in Private Primary Schools, Educated Section and Percentage of people earning with the Wealth Estimate, imply that setting up an ATM in an affluent area augments transactions. It has also been observed that people living away from their family are likely to visit ATMs more frequently which is corroborated by the features Percentage of single people, Percentage of people single with roommates etc. Features like ethnicity or race, showed a very small correlation with the objective function and hence were not considered significant. The observations show that the features obtained are logically sound and it can be concluded that our model performs efficiently.

### 6.2 Scoring

The task involves adding a name-tag classification for each ATM as discussed in the data pre-processing stages. We calculated the average score for each ATM network in each unique zipcode and the frequency of the ATM network in the zipcode according to the equation.

$$S_{atm} = S_{zipcode} \times (ASA) \times Frequency \quad (4)$$

where, $S_{atm}$ = Score of ATM network in a particular zipcode, $S_{zipcode}$ = score of the zipcode and $ASA$ = Average Score of ATM network. The score of an ATM network in a given county is given by:

$$S_{atm(global)} = \sum_{zip \in county} S_{zip} \times (ASA|_{zip}) \times Frequency|_{zip} \quad (5)$$

The score we obtain from the above equation is the result from the global model. In the local model, the data is divided into the number of counties. In each county, 7 clusters are initially set in order to visualize the important patterns in the data. From the clustered data we reverse engineer the features using the Random Forest classifier. The most important 20 features are chosen per count and weighted sum is computed.

$$S_{atm(local)} = WT_{rf}^T \times x \times (ASA|_{county}) \quad (6)$$

where, $W_{rf}$ = weight vector obtained from random forest classifier, $x_i$ = feature data In this method we directly compute the score of the county. The name-tag weightage of each bank network is computed per county and multiplied with the county score in order to get the ATM network score in the given county. The score of each bank gives us an overview of the relative revenue advantage of an ATM network. Finally, to compute the overall score, we compute a linear fusion of both global and local features $S_{total}$:

$$S_{atm} = (1 - \alpha) S_{county(local)} + \alpha S_{county(global)} \quad (7)$$

In the above equation, $\alpha$ is a variable which controls how much the global feature is dominant in the prediction model. The $\alpha$ value may vary depending on the dataset, in our case $\alpha = 0.35$, as we intuitively expect the local features to play a larger role in determining the overall score.

### 6.3 Validation

Validation of the scores obtained was attempted on the basis of publicly available data. However, it was noted that the publicly available data were all derivatives of the actually required validation data. Hence, the technique used was to validate on data which seemed to correlate with the ATM distribution data such as spending maps in US counties [8], deposits made at the financial institutions of an area [4] etc. Even though these data may not be a strong indicator of validation but should at least provide weak indicators of the success of the scoring methodology. Actual ATM transactions distribution data was not available publicly and hence could not be presented for validation purposes.

## 7 CONCLUSION

Thus, we would like to propose the above system, and conclude that the hybrid approach is ideologically optimal. The important aspect that the model takes into consideration is that the feature weights varies with county location. We observe that the factors with higher weights vary greatly with the counties. So it would not be very prudent to generalize over all the counties with a specific set of factors. That is why taking the weighted score summation of two models, one which considers local factors and the other which considers the global trends would be the most generalized way to go about solving this problem. The reason why we took into account both the global and local models instead of considering only the local model is that it would otherwise lead to over-fitting, and on the other hand, without local data the model would prove to be very generic. The weight of the local region in a county is kept higher than the global weight as the significant local factor has to be given more importance. The weight of local model is to be kept as 0.65 and 0.35 assigned to the global model. From the score estimate of the county regions, we can maximize our revenue if there is a venture for opening a new ATM network. As we already have the score of individual revenues county wise. We can keep in the cost of setting up a new ATM in a particular region as a limiting factor. Then we can optimize the function of the reward (taken as county score) and the penalty (cost of setting up an ATM) over a few counties. Then we can decide which public spots we can set up ATMs in order to reach maximum revenue.